\documentclass[twocolumn]{aastex62}
\renewcommand{\sun}{$_\odot$}

\begin{document}

\title{A Case Study of On-the-Fly Wide-Field Radio Imaging Applied to the \\Gravitational-wave Event GW\,151226}

\correspondingauthor{K. P. Mooley}
\email{kmooley@caltech.edu}

\author{K. P. Mooley}
\altaffiliation{Currently Jansky Fellow at NRAO, Caltech.} 
\affil{Denys Wilkinson Building, Keble Road, Oxford OX1 3RH, UK.}
\affil{NRAO, P.O. Box O, Socorro, NM 87801, USA}
\affil{Caltech, 1200 E. California Blvd. MC 249-17, Pasadena, CA 91125, USA}

\author{D. A. Frail}
\affil{NRAO, P.O. Box O, Socorro, NM 87801, USA}

\author{S. T. Myers}
\affil{NRAO, P.O. Box O, Socorro, NM 87801, USA}

\author{S. R. Kulkarni}
\affil{Caltech, 1200 E. California Blvd. MC 249-17, Pasadena, CA 91125, USA}

\author{K. Hotokezaka}
\affil{Department Astrophysical Sciences, Princeton University, Peyton Hall, Princeton, NJ 08544, USA}

\author{L. P. Singer}
\affil{Astrophysics Science Division, NASA Goddard Space Flight Center, Mail Code 661, Greenbelt, MD 20771, USA}

\author{A. Horesh}
\affil{Racah Institute of Physics, Hebrew University of Jerusalem, Israel}

\author{M. M. Kasliwal}
\affil{Caltech, 1200 E. California Blvd. MC 249-17, Pasadena, CA 91125, USA}

\author{S. B. Cenko}
\affil{Astrophysics Science Division, NASA Goddard Space Flight Center, Mail Code 661, Greenbelt, MD 20771, USA}
\affil{Joint Space-Science Institute, University of Maryland, College Park, MD 20742, USA}

\author{G. Hallinan}
\affil{Caltech, 1200 E. California Blvd. MC 249-17, Pasadena, CA 91125, USA}

\begin{abstract}

We apply a newly-developed On-the-Fly mosaicing technique on the NSF's Karl G. Jansky Very Large Array (VLA) at 3 GHz in order to carry out a sensitive search 
for an afterglow from the Advanced LIGO binary black hole merger event GW\,151226. 
In three epochs between 1.5 and 6 months post-merger we observed a 100 deg$^2$ region, with more than 80\% of the survey region
having a RMS sensitivity of better than 150 $\mu$Jy/beam, in the northern hemisphere having a merger containment probability of 10\%. 
The data were processed in near-real-time, and analyzed to search for transients and variables.  
No transients were found but we have demonstrated the ability to conduct blind searches in a time-frequency phase space where the predicted afterglow signals are strongest. 
If the gravitational wave event is contained within our survey region, the upper limit on any late-time radio afterglow from the merger event at an assumed mean distance of 440 Mpc is about $10^{29}$ erg s$^{-1}$ Hz$^{-1}$. 
Approximately 1.5\% of the radio sources in the field showed variability at a level of 30\%, and can be attributed to normal activity from active galactic nuclei. 
The low rate of false positives in the radio sky suggests that wide-field imaging searches at a few Gigahertz can be an efficient and competitive search strategy. 
We discuss our search method in the context of the recent afterglow detection from GW\,170817 and radio follow-up in future gravitational wave observing runs.

\end{abstract}


%
%
%
%

\section{Introduction}\label{sec:intro}

The era of gravitational wave astronomy has begun. 
The Advanced Laser Interferometer Gravitational-wave Observatory \citep[aLIGO;][]{abbott2016-aLIGO} first observing run (O1), which ran in the last quarter of 2015, 
reported two significant binary black hole (BBH) merger events GW\,150914 and GW\,151226 \citep{abbott2016-gw150914,abbott-gw151226}.  
The second observing run of aLIGO (O2) began in late 2016 and ended on 25 August 2017, with the Virgo detector joining on 01 August 2017 to form a three detector network. 
Two more significant BBH mergers were detected, GW\,170104 and GW\,170814 \citep{abbott-gw170104,abbott-gw170814}, and for the first time gravitational waves were detected from the coalesecence of two neutron stars GW\,170817 \citep{abbott-gw170817}. The notification of the discovery of BBHs triggered a world-wide, panchromatic search for their electromagnetic counterparts \citep[EM; e.g.][]{abbott2016-gw150914-followup,abbott2016-gw150914-followup-supplement,ekp+16,cbs+16,csp+16,gbs+17,rbg+17}.  
Thus far no conclusive time-variable or quiescent emission has been found at any wavelength for BBHs. 
In contrast, the binary neutron star merger GW\,170817 was accompanied by EM signals detected at all wavelengths, including prompt gamma-ray emission \citep{fermi-gw170817}, 
fast-fading optical/NIR \citep{swope-gw170817}, and delayed X-ray \citep{xray-gw170817} and radio emission \citep{radio-gw170817,mooley2017-gw170817}.


Prior to the detection of the EM counterpart to GW\,170817, radio emission was widely expected to arise on a wide range of timescales and luminosities from compact binary star mergers. 
Mergers involving neutron stars leave behind significant neutron-rich debris that settles into a disk.
In the conventional picture, most of the debris disk is accreted by the newly formed black hole post-merger (leading to a short GRB in the case of binary neutron stars) and a small amount, about 0.01M\sun, is 
ejected \citep[e.g.][]{rosswog1999,bauswein2013,hotokezaka2013,rosswog2013,radice2016}. 
The forward shock into the ISM swept up by the (sub-relativistic) ejecta is expected to generate broad-band synchrotron emission. 
This gives rise to a milliJansky-level radio transient on timescales of months to years \citep[e.g.][]{nakar2011,hotokezaka2016}, peaking at frequencies around a few Gigahertz. 
In cases where relativistic jets are formed and beamed away from the observer, the deceleration of the jet through interaction with the ISM eventually opens up the emission cone into the 
observer's line of sight. 
Such orphan afterglows appear as radio transients on timescales of weeks to months \citep{hotokezaka2015,hotokezaka2016}. 
This simple jet model likely needs to be modified as the X-ray and radio emission from GW\,170817 are best understood as the breakout of wide-angle, mildly-relativistic outflow (consistent with a ``cocoon'') 
of material entrained by the jet \citep{ldmw17,kns+17,gnph17,mooley2017-gw170817,ruan2017-gw170817}.

For neutron star mergers, there is another possible channel for generating radio emission. 
A millisecond magnetar is a plausible merger remnant, where the magnetar wind drives a strong shock into the ejecta and the reverse shock results in a ``plerion'' (cf. the Crab nebula). 
A strong plerionic radio emission, which is isotropic and independent of the ambient density, is expected on timescales of few months \citep{piro2013}.  
There is also a strong radio signal expected at late times ($\sim$year timescale; even at small ISM densities) since the magnetar can drive the ejecta to relativistic velocities \citep[e.g.][]{mtf+17}. 
Past searches for late-time radio emission from short GRBs have put constraints on the phase space of kinetic energy, ejecta mass, and ISM densities, and on the magnetar scenario \cite{metzger2014,horesh2016,fong2016}.

In the case of binary black holes, radio emission is not widely expected to arise from baryonic poor environments. 
However, if the BBH merger launches a relativistic jet into a dense, gas-rich environment radio emission is expected at a level of order 10--100 $\mu$Jy (at an assumed distance of $\sim$400 Mpc), 
on timescales of 10$^5$ s and at frequencies around a few GHz \citep{yao16,khm17}.


There are several challenges that must be overcome in any observational effort designed to detect EM counterparts. 
At radio wavelengths the main barriers are achieving the necessary sub-mJy sensitivity over the large aLIGO error regions of 100's of deg$^2$ at GHz frequencies needed to optimally test existing theoretical models. 
Equally important is having the ability to rapidly identify and reject any variable sources that could lead to misidentification, and to provide sub-arcsecond localization for any viable candidates \citep{metzger2012,mwb15,hotokezaka2016}. 
In the face of such challenges several alternative strategies have been adopted. 
\citet{pck+16} and \citet{bhalerao-atlas} have taken the approach of carrying out radio follow-up of optically-selected candidates, identified from wide-field imaging surveys such as the 
Dark Energy Camera \citep[DECam;][]{cbs+16} or the Palomar Transient Factory \citep[PTF;][]{law2009}. 
This approach avoids imaging large sky areas and it can make deep observations over a wide frequency range. 
Another approach has been to shift to lower frequencies in order to utilize existing wide-field instruments. 
Radio searches spanning 50--100 deg$^2$ were carried out with the Australian Square Kilometre Array Pathfinder \citep[ASKAP;][]{bannister2015}, the Low-Frequency Array  \citep[LOFAR;][]{broderick2015,broderick2016}, 
and the Murchison Widefield Array \citep[MWA;][]{kmr+16} telescopes at frequencies of 863 MHz, 145 MHz, and 118/154 MHz, respectively. 
While these surveys did not have the sensitivity and angular resolution needed to detect and provide arcsecond-localization of any putative radio afterglows of the GW events, 
they were good proof-of-concept experiments for the follow up of aLIGO sources using wide-field blind radio observations.

\begin{figure*}[!t]
\centering
\includegraphics[width=3.5in,viewport=20 35 410 320,clip]{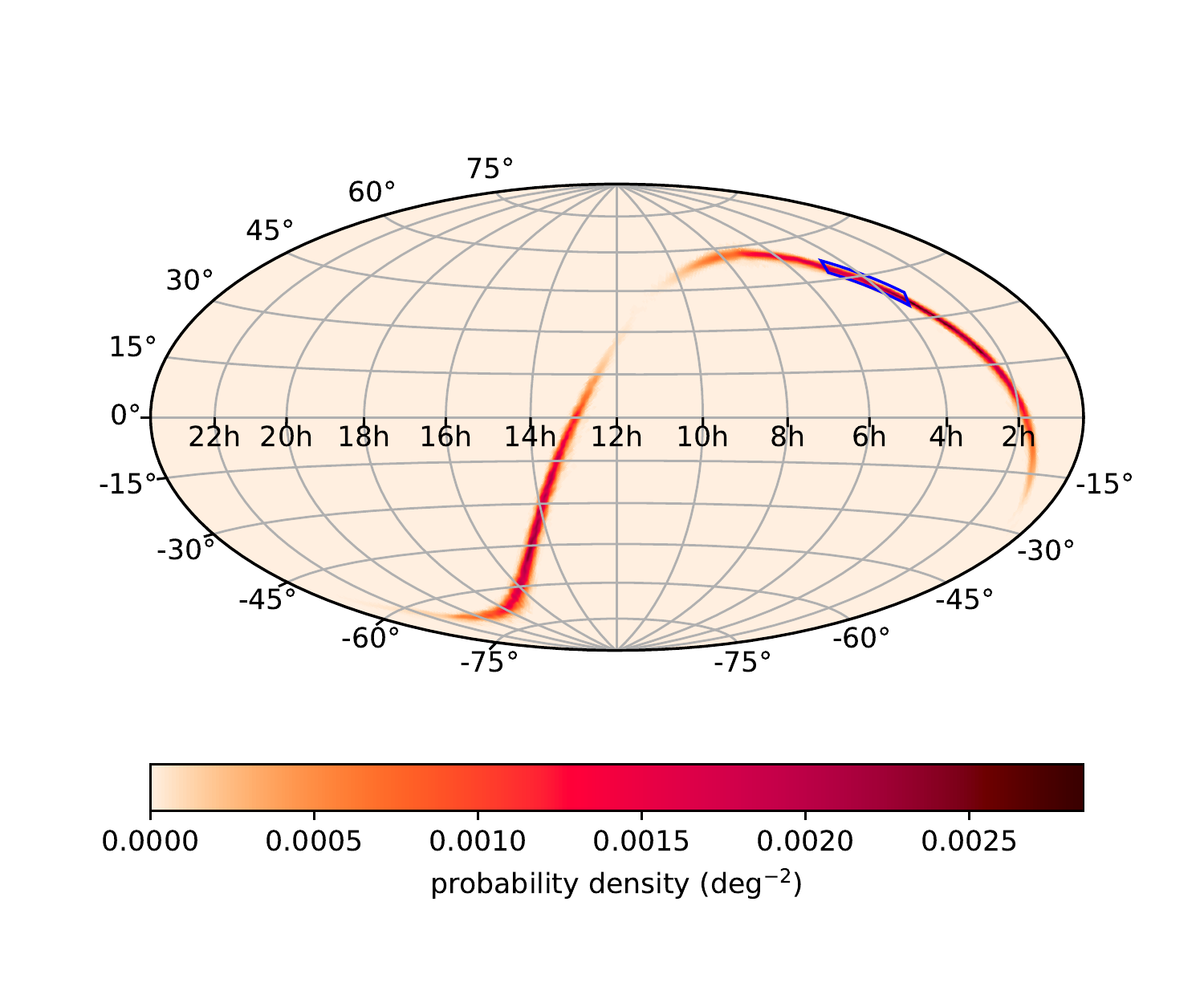}
\includegraphics[width=3.5in]{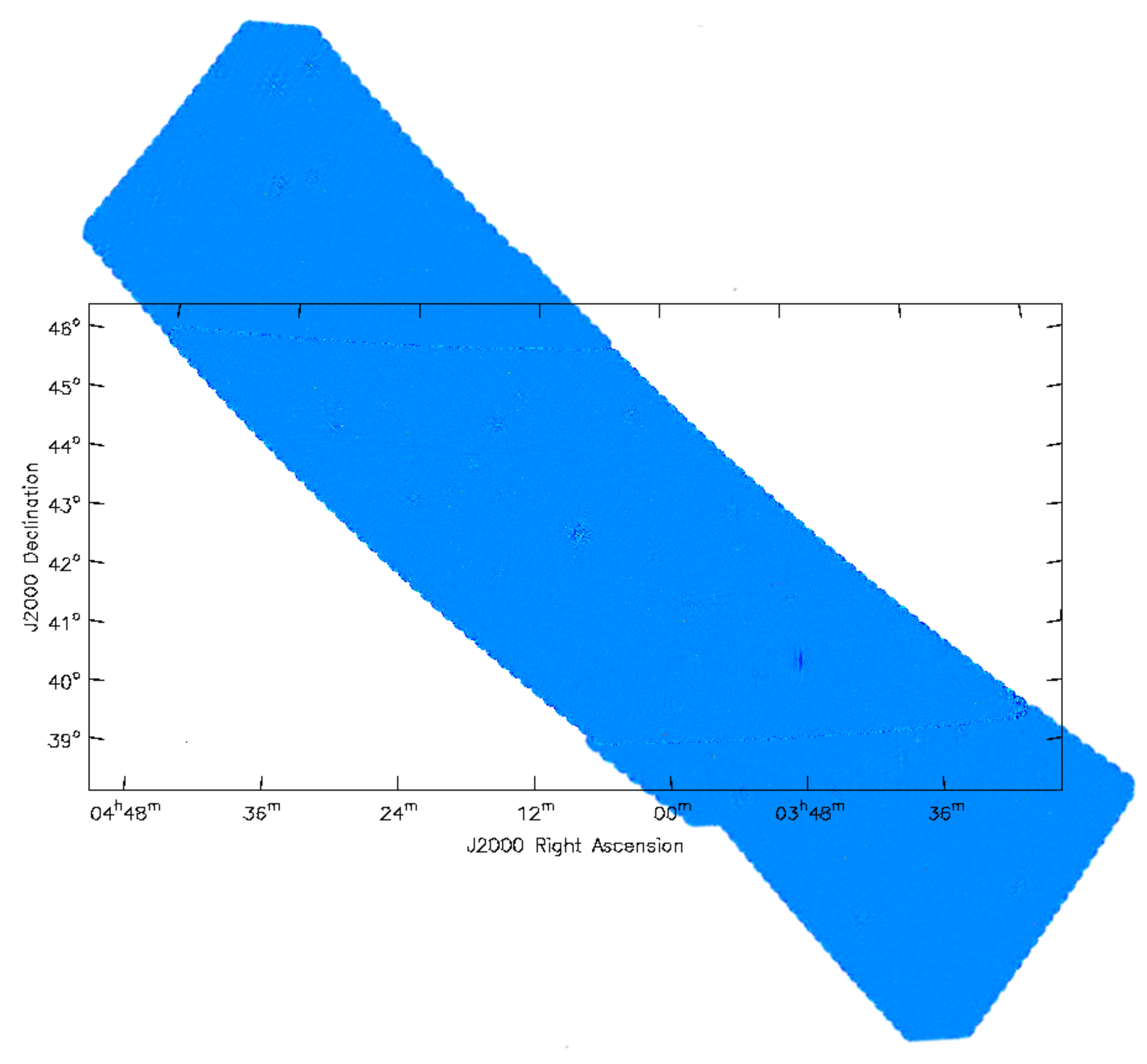}
\caption{Left: The LALInference localization of GW\,151226. 
The 90\% credible region is 1240 deg$^2$. 
The orange color scale represents the containment probability (black for maximum probability and white for the least).
The 100 deg$^2$ region observed with the Jansky VLA, having a containment probability of 10\%, is outlined in blue.
Right: Our radio image mosaic of this region from the first observing epoch. 
The colorbar runs from 100 $\mu$Jy (blue) to 1500 $\mu$Jy (white).
The median image noise is $\sim$110 $\mu$Jy/bm image noise. 
The noise is higher around bright sources and edges of the survey region.}
\label{fig:gw151226_aitoff}
\end{figure*}

In this paper we take a more direct approach by using the VLA On-the-Fly mosaicing capability to make sensitive, wide-field observations at arcsecond resolution and at Gigahertz frequencies. 
We use this new mode to carry out follow-up observations of GW\,151226. 
In \S\ref{sec:obs_proc} we describe the VLA observations and the data reduction carried over a 100 deg$^2$ region. 
The search for transients and the identification of variables is described in \S\ref{sec:var_tran}. 
This search is discussed in the context of future aLIGO-Virgo observing runs and radio follow up programs in \S\ref{sec:discussion}.

\section{Observations and Data Processing}\label{sec:obs_proc}

\subsection{The Gravitational Wave Event GW~151226}

The GW\,151226 gravitational wave signal was initially identified by the GstLAL compact binary coalescence search \citep{messick2017} 
of the data from the LIGO Hanford and Livingston detectors at 2015 Dec 26.15, and localized by the BAYESTAR code \citep{singer2016}, which is sensitive to compact binary star coalescence events.
The false alarm rate for this event was reported as being lower than one per hundred years \citep{unh+16,messick2017}. 
Using Bayesian Markov-chain Monte Carlo and nested sampling to perform forward modeling of the full GW signal including spin precession and regression of systematic 
calibration errors \citep[LALInference;][]{veitch2015} the event was localized to within 1240 deg$^2$ (90\% credible region), significantly improving over the BAYESTAR localization. 
The LALInference sky map of the gravitational wave event, together with our VLA survey region, is shown in the left panel of Figure~\ref{fig:gw151226_aitoff}. 

GW\,151226 marks the second direct detection of gravitational waves. 
Detailed offline analysis of the aLIGO data recovered the gravitational wave signal with a significance greater than 5$\sigma$. 
The initial (individual) and final black hole masses were estimated to lie between 5--22 M\sun and 19--27 M\sun, respectively. 
The luminosity distance was estimated to lie between 250--620 Mpc, i.e. $0.05<z<0.12$ \citep[the ranges represent the 90\% credible interval;][]{prl_gw151226}.

\subsection{On-the-Fly mosaicing}\label{otfm}

The recent refurbishment of the VLA has increased its instantaneous sensitivity by almost an order of magnitude \citep{pnj+09} but its field of view (several arcmin FWHM) at GHz frequencies is still relatively small. 
As a result, in the conventional pointed observing mode the slew-and-settle time of the antennas can become a significant fraction of the on-sky integration time, especially for wide-field imaging. 
These overheads can be minimized through the use of On-the-Fly mosaicing (OTFM), where the antennas are driven at a non-sidereal rate and visibilities are recorded continuously, to significantly 
improve the efficiency of wide-field surveys. 
OTFM is therefore naturally the observing mode of choice for blind transient searches and LIGO follow up observations, both of which require observations over wide fields of view and over multiple epochs. 
One additional advantage of the OTFM method over wide-field radio telescopes is that the VLA can be made to image irregularly shaped GW error regions, with no loss in sensitivity.

The OTFM observing mode has been recently commissioned (Mooley et al., in prep.) on the VLA, and this mode can increase the observing efficiency by $\gtrsim$10\% for surveys not requiring 
high sensitivity (10s-100s of $\mu$Jy RMS noise is sufficient). 
In the VLA implementation of OTFM, the antennas slew with a uniform speed along a long ``stripe'' usually in constant Right Ascension or Declination. 
The antennas are then stepped in Declination or Right Ascension to the next stripe and so on, in order to observe the survey region in a ``basket-weave'' pattern of the antenna motion.

\begin{table*}[htp]
\centering
\scriptsize
\caption{Observing Log}
\label{tab:observations}
\begin{tabular}{lllllllll}
\hline\hline 
No. & Start Date  & Reg/Epo & LST         & $\Delta$t & Array  & RMS  & Beam & Phase calibrator sources \\
    & (UT)        &         & (h)         & (days) & Config.& ($\mu$Jy/bm)  & (``)  \\
\hline
1   & 11 Feb 2016 & R2E1    & 00.0--03.8  & 47 & C    & \ldots   & 8    & J0348+3353, J0414+3418, J0438+4848\\
2   & 14 Feb 2016 & R1E1    & 22.5--02.3  & 50 & C    & 115      & 8    & J0354+4643, J0438+4848, J0439+4609\\
3   & 05 Apr 2016 & R1E2    & 06.0--09.8  & 101 & C    & \ldots   & 8    & J0354+4643, J0438+4848, J0439+4609\\
4   & 05 Apr 2016 & R2E2    & 05.0--08.8  & 101 & C    & 112      & 8    & J0348+3353, J0414+3418, J0438+4848\\
5   & 27 Jun 2016 & R1E3    & 22.6--02.4  & 184 & B    & \ldots   & 3    & J0354+4643, J0438+4848, J0439+4609\\
6   & 30 Jun 2016 & R2E3    & 06.1--09.9  & 187 & B    & 150      & 3    & J0348+3353, J0414+3418, J0438+4848\\
\hline
\multicolumn{9}{p{5.6in}}{Notes: (1) Entries from left to right include the observing run (No.), the start date, the survey region and epoch (Reg/Epo), the start and stop LST time, 
the time in days since the GW event, the VLA array configuration, the RMS noise for each epoch, the synthesized beam size and a list of phase calibrators used for each epoch.
(2) RMS refers to the 50th percentile of the RMS noise across the survey region for the given epoch.}
\end{tabular}
\end{table*}

\subsection{Radio Observations}

Using the aLIGO LALInference sky localization map for GW\,151226, we selected a 100 deg$^2$ maximum-probability region\footnote{Bounding 
coordinates are ($\alpha$=50.25$^o$,$\delta$=38.25$^o$), 
(54.25,34.30), (70.00,51.50), (74.50,47.50). The survey area of 100 deg$^2$ was motivated 
by our minimum sensitivity requirement, $\sim$100 $\mu$Jy per epoch, in order to catch any putative radio afterglow, keeping in consideration also the 
allotted VLA time for our observations. Furthermore, choosing additional 50 deg$^2$ of high-probability area increased the containment probability only 
by 0.1\%, therefore we did not increase the survey area beyond 100 deg$^2$.}
in the Northern hemisphere (see Figure~\ref{fig:gw151226_aitoff}), having a containment probability of 10\%, for follow up. 
Observations were carried out\footnote{Under project code VLA/16A-237} across three epochs (E1, E2, E3) with the NSF's Karl G. Jansky Very Large Array (VLA) in the B and C array configurations. 
S-band (2--4 GHz) was chosen to maximize survey speed and catch a putative late-time afterglow. 
To maximize the continuum imaging sensitivity, we used the Wideband Interferometric Digital Architecture (WIDAR) correlator with 16 spectral windows, 
64 2-MHz-wide channels each to get 2 GHz of total bandwidth centered on 3.0 GHz.
Two basebands, centered on 2.5 GHz and 3.5 GHz, consisted of 8 spectral windows each.
We used the OTFM mode (\S\ref{otfm}) and used 1-sec integrations to minimize the amplitude smearing. 
A log of the observations is given in Table~\ref{tab:observations}.

In order to facilitate the scheduling, we divided the survey area into two regions, R1 (Dec $39.5^o$--$46.0^o$) and R2 (Dec $34.2^o$--$39.5^o$ and $46.0^o$--$51.5^o$), 
which were observed in each epoch within a span of a few days. 
Our dynamically-scheduled observing blocks (for each region and each epoch) were $\sim$3.75 hours and were designed using {\tt OTFSim} (Mooley et al. 2017, in prep.). 
Given these boundaries of the survey region, OTFSim automatically selected the appropriate path of antennas and complex gain calibrators to minimize the slew time.
The antenna slewing was done along constant declination in order to ensure uniform coverage and sensitivity over the survey region. 
Thus, we designed our OTFM observations to slew the antennas purely in right ascension at a rate of 2 arcmin per second (on-sky rate of 1.6 arcmin per second), 
stepping northwards by 10.6 arcmin (FWHM of primary beam divided by $\sqrt{2}$, to get approximately uniform RMS noise across the survey region) after each slew, in a basket-weave fashion. 
The correlator phase center was stepped every 4 arcmin in right ascension (every 2 s) to ensure that the antenna slew during each scan in the observation was well within one full primary beam.
3C147 was used as the flux density and bandpass calibrator.


\begin{figure}
\centering
\includegraphics[width=3.5in]{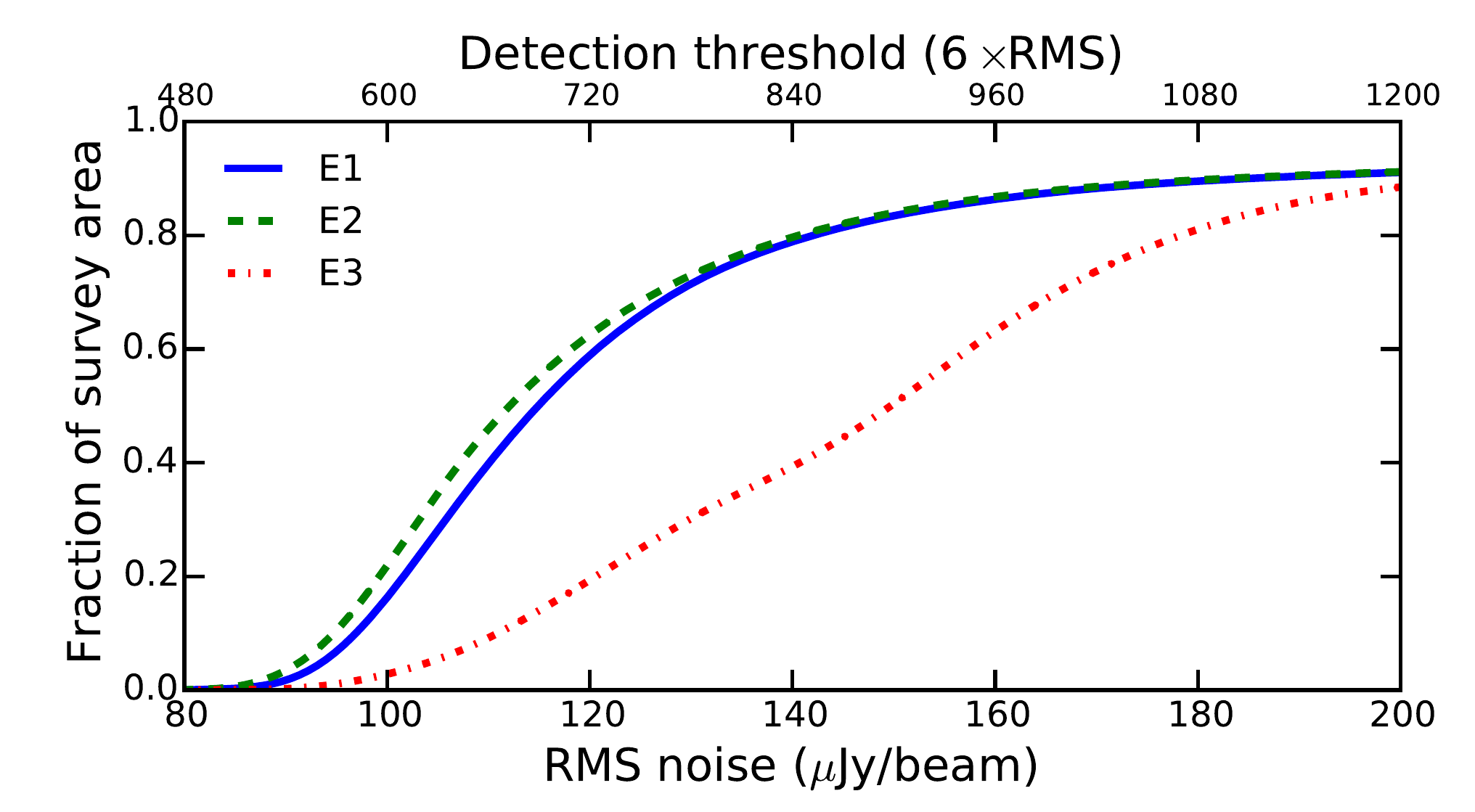}
\caption{Cumulative plots of the RMS noise across the survey region for the three epochs of observations reported here.
The source detection threshold (6$\sigma$) is shown on the upper x-axis.
About 10\% of the survey region, corresponding to areas around bright sources and field edges, have RMS noise $>$200 $\mu$Jy/beam.}
\label{fig:rms_cumulative}
\end{figure}

\subsection{RFI flagging, Calibration, Imaging and Source Finding}\label{sec:obs_proc.cal}

Immediately after the observations for each epoch/region (each observing block) were complete, we downloaded the raw data from the VLA archive, onto the lustre file system at the NRAO AOC in Socorro. 
The data were then calibrated using the NRAO scripted pipeline (in CASA 4.5.0) with quacking\footnote{Quacking is the process of removing (potentially bad) data 
from the beginning and end of each scan.} 
of the target fields removed. 
Due to substantial satellite-induced RFI, the data between 2.12--2.37 GHz (the second and third spectral windows) were fully flagged for all observations.
Single scans (5500--5700 scans for each region per epoch) were then split from the calibrated dataset into individual measurement sets. 
The calibration and splitting processes for each observing block took about 36 hours.

We then ran a distributed imaging process on 70 cores across 5 computer nodes at the NRAO AOC using CASA task {\it clean}, such that each core 
was imaging a single scan at any given time. 
To enable linear mosaicing, we convolved the images with fixed circular synthesized beams\footnote{The synthesized beam sizes found by 
{\it clean} ranged between $\sim$7\arcsec--9\arcsec~for C config data and between $\sim$2\arcsec--3\arcsec~for B config data. }
of 8 arcsec (for the C config observations) or 3 arcsec (for the B config observations).
The imaging of each scan was itself done over two iterations.
In the first iteration we cleaned using Briggs weighting and a robust parameter of zero in order to minimize imaging artifacts.
Images with sources brighter than 20 mJy were then used to derive and apply a single phase self-calibration procedure. 
Clean boxes were then derived based on the sources found using the {\tt ia.findsources} tool in CASA. 
This was followed by clipping of the visibilities in amplitude (in all spectral windows; in order to minimize the effect of RFI) at a level equal to the sum of the mean and three 
times the standard deviation of the visibility amplitudes in the spectral window with the least RFI. 
The second imaging iteration used the self-calibrated visibilities to image using natural weighting and clean boxes. 
During each cleaning step we used 700 clean iterations, two Taylor terms, cyclefactor parameter of 7, and a clean stopping threshold of 0.3 mJy.
The pixel size was so chosen as to sample the synthesized beam across at least four pixels.
The images had a center frequency of 3.0 GHz, except for the observation on 27 June 2017 (where the data had substantial RFI between 2--3 GHz) for which the center frequency 
is 3.3 GHz\footnote{For sources with steep spectral indices of $\pm$1, the flux densities between the 3 GHz images and the 3.3 GHz images will differ by $\sim$10\%. 
This biases our variability search slightly in favor of finding steep spectral index sources, as discussed in more detail in \S\ref{sec:var_tran}.}. 
A small fraction of the single-scan images had strong spike-shaped artifacts (known bug in CASA {\tt clean} when using two or more Taylor terms; currently being fixed), and 
for these scans we chose the Briggs-weighted images (output during the first imaging iteration described above) instead of the ones with two Taylor terms.
Based on our inspection of the single-scan images, we expect that the Briggs weighting reduces the integrated flux density of extended sources in the survey region 
by a fraction less than or equal to $\sim$10\%. For unresolved sources there will be no change in flux density due to the introduction of the Briggs-weighted images.

Linear mosaicing of the single-scan images was then carried out using {\tt FLATN} in AIPS, followed by cropping of the mosaic into 4096$\times$4096 pix$^2$ sub-images.
We used the primary beam parameters from Perley et al. (EVLA Memo 195) during the linear mosaicing step. The imaging and linear mosaicing processes together required about 6 hours per region/epoch. 
We then made a background noise map for each sub-image using {\tt RMSD} in AIPS, which was then supplied to {\tt SAD} for the 
cataloging of sources down to\footnote{For the B config data there are $2\times10^8$ synthesized beams across the 100 deg$^2$ survey region. 
So the 6$\sigma$ detection threshold ensures $<$1 false detection due to noise (assuming Gaussian statistics).} 6$\sigma$. 
The cumulative noise plots made from the background RMS noise maps is shown in Figure~\ref{fig:rms_cumulative}. 

As we are interested in only unresolved and partially resolved sources, we chose only those sources from each epoch that had an integrated-to-peak flux density ratio of $<$1.5 and prepared a merged catalog, consisting of 5307 sources. 
For sources in the merged catalog that were not detected in some of the epochs, we set the peak flux density to be equal to the peak pixel value at the location of the source. 
For those sources having the integrated flux density lower than the peak flux density, we set the integrated flux density to be equal to the latter. 
This cataloging process required about 1 hour. 
The total time required from downloading the raw data for each observing block and producing a final source catalog was 43 hrs.

\section{Transient and Variability Search}\label{sec:var_tran}

We used the catalog of 5307 sources from \S\ref{sec:obs_proc.cal} to carry out a search for transients sources which appeared 
or disappeared in one or more of the three epochs. 
No transients were found to a 6$\sigma$ limit of 670 $\mu$Jy (50\% completeness threshold for the merged catalog over 3 epochs and 100 sq deg).

The same catalog was used to investigate variability following a process described in more detail in \citet{mooley2016}. 
In short, for each source we calculate the variability statistic, $V_s=\Delta S/\sigma$ for each epoch and a modulation index $m=\Delta{\rm S}/\bar{\rm S}$, where S is the flux density, 
$\bar{\rm S}$ is the average flux density, $\Delta$S is the flux density difference, and $\sigma$ is the RMS noise. 
Significant variables were identified as those sources having $V_s$ larger than 4$\sigma$, and the absolute value of the modulation index, $|m|$, larger 
than 0.26 (i.e. a fractional variability, $f_{\rm var}>0.3$). 
The constraint on $V_s$ ensures fewer than one false positive will be detected as a variable source in our search, assuming Gaussian statistics. 
Considering that our flux scale is accurate to only 3--5\% \citep{thyagarajan2011,mooley2013}, the artificial variability induced on account of our 
imaging parameters \citep[$\sim$10\%; see \S\ref{sec:obs_proc.cal} and also ][]{mooley2016}, and our usage of the true primary beam instead of the smeared OTFM beam, 
the constraint on the fractional variability is needed to minimize the number of false positives.

We used epoch E1 as the reference epoch and compared E2 and E3 independently with this epoch using the $V_s$ and $m$ for finding significant variable sources.
Based on our inspection of the ratios of source flux densities between E2 and E1, we had to apply a multiplicative correction factor of 1.07 to the source flux densities in E2.
Similarly, our comparison of E3 versus E1 instructed us to apply a multiplicative correction factor\footnote{This factor is rather high, but is partly due to angular resolution difference between the two epochs.
The flux lost in E3 is due to the missing short spacings in the VLA B array configuration, and could not be recovered to the C array flux even after imaging with a restricted UV range and UV tapering.
In principle the missing short spacings should not affect true point sources, but the robust comparison of flux densities across different array configurations is an issue that needs to be investigated further.} 
of 1.30 to the source flux densities in E3.
This correction factor made the distribution of the variability statistic $V_s$ symmetrical, but also broadened it to a Gaussian-like distribution ($\sigma=2.5$, hence our $4\sigma$ variability selection criterion in this case is $V_s>10$) 
rather than a Student-t distribution \citep[cf. Figure 9 of ][]{mooley2016}.
During the variability search between E3 and E1, we also discovered that a significant fraction of the sources in the final source catalog were resolved out in E3, and abnormally large number of variables appeared below $\sim$1 mJy 
(this can again be attributed largely to the angular resolution differences).
Hence, for the comparison between E3 and E1 we restricted our search to only those sources (total of 2782) having integrated-to-peak flux density ratio\footnote{This constraint somewhat 
reduces our ability to reliably find variable sources below $\sim$1 mJy, but at the same time significantly mitigates false positives.} of $<$1.2.

\begin{figure}
\centering
\includegraphics[width=3.5in]{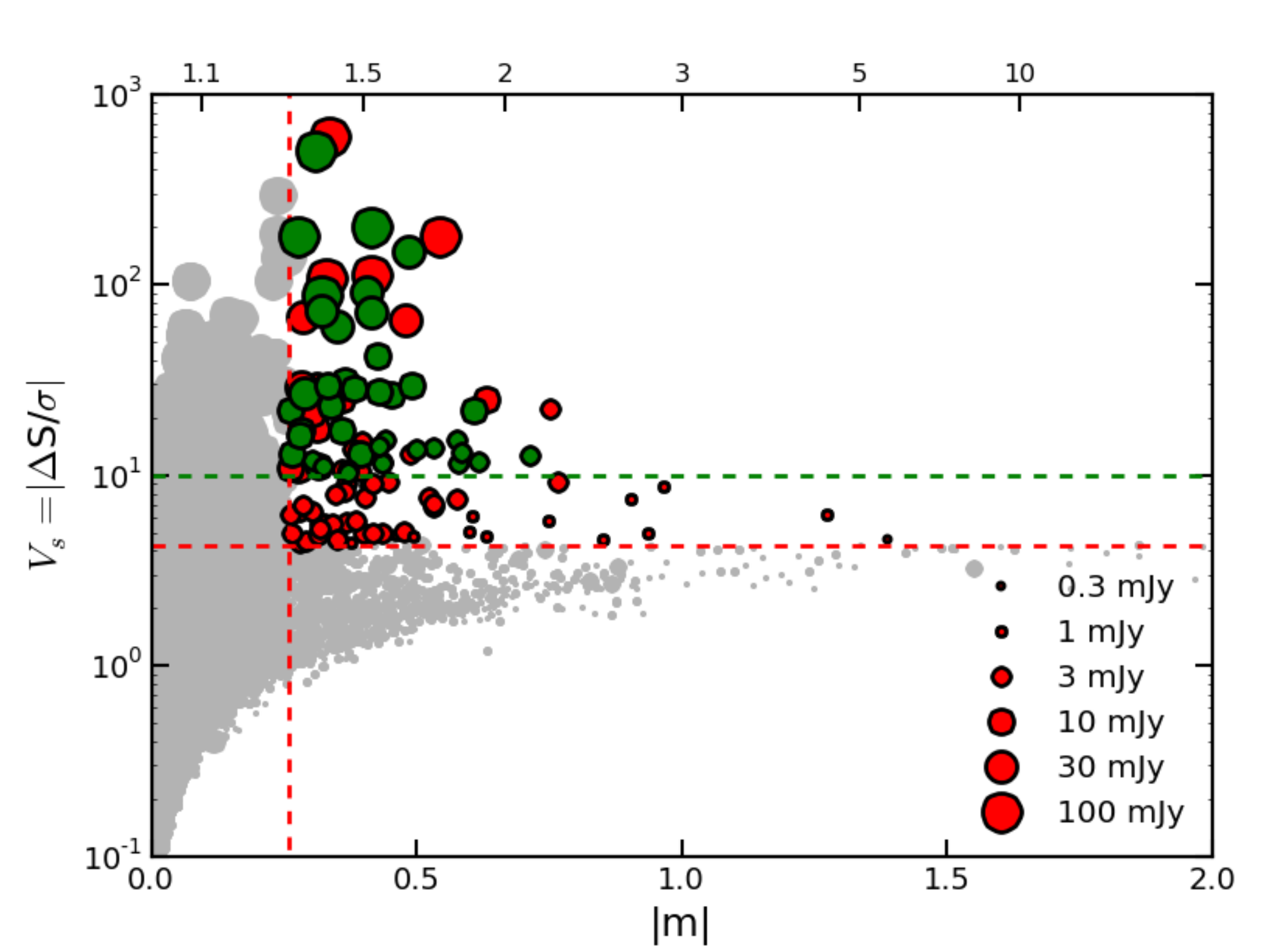}
\caption{Variability statistic ($V_s$) versus the modulation index ($m$) for the unresolved/partially-resolved sources in our merged catalog. 
Grey points indicate sources (from the E1-E2 and E1-E3 comparisons) that are not significant variables.
The red points are the selected variables between E1 and E2 (totaling 72 out of 5307 sources), and green 
points are those between E1 and E3 (totaling 42 out of 2782 sources).
The horizontal red and green dashed lines indicates the variability selection criteria in $V_s$ for these two respective cases.
The vertical red dashed line is the variability selection criteria in modulation index.
The flux densities of the sources defines the marker size.
Top horizontal scale is the fractional variability $f_{\rm var}$, defined as the ratio of the flux densities of a source between two epochs.
See \S\ref{sec:var_tran} for details.}
\label{fig:var}
\end{figure}

The plot of the variability statistic versus the modulation index is shown in Figure~\ref{fig:var}. 
We found 72 significant variable sources between E1 and E2 and 42 variables between E1 and E3, having fractional variability larger than 30\%. 
This indicates a variability fraction of $1.5\pm0.2$\% for timescales between few weeks and few months.

\section{Discussion and Future Prospects}\label{sec:discussion}

\begin{figure}
\centering
\includegraphics[width=3.5in]{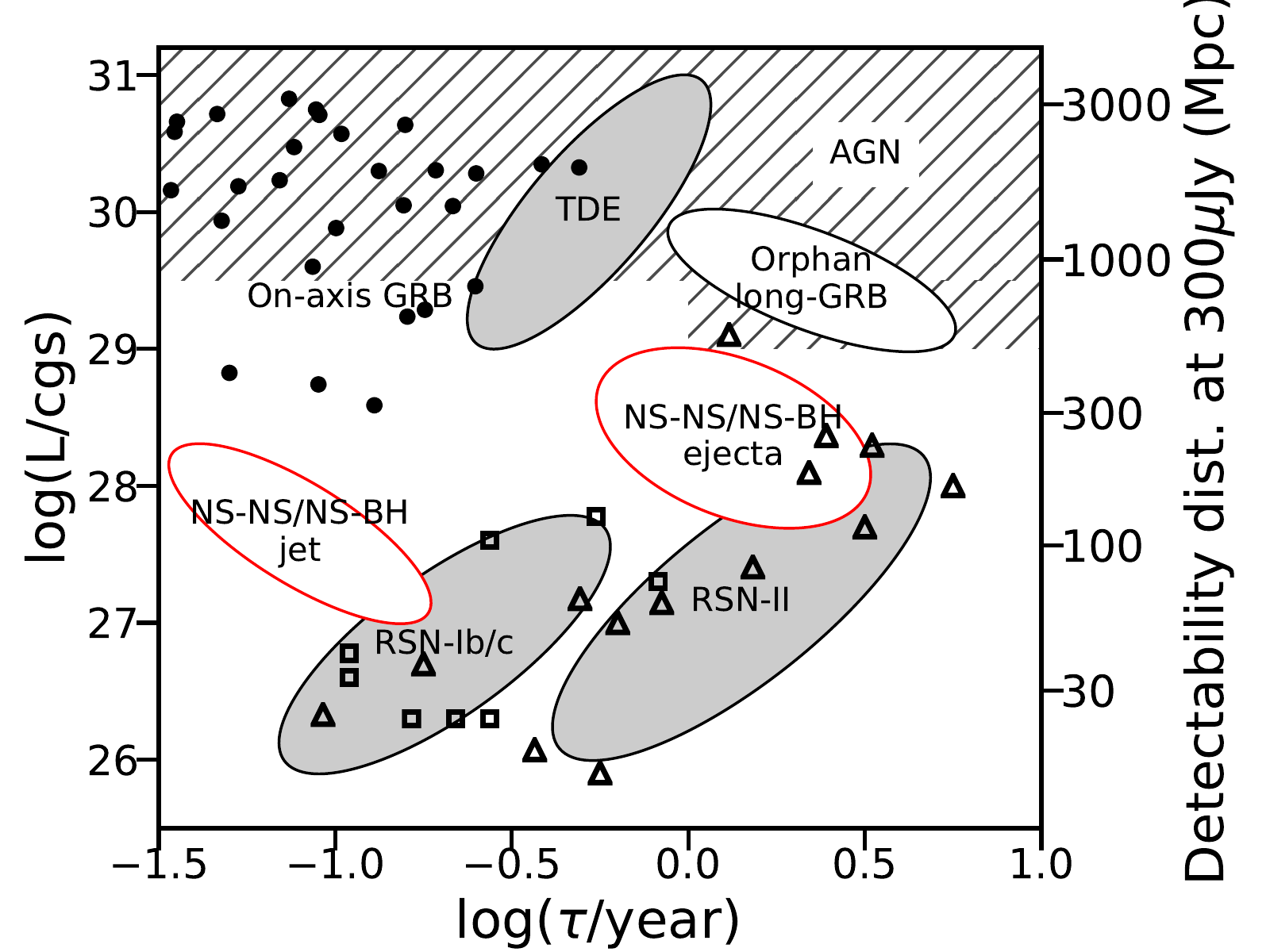}
\caption{The luminosity (in erg s$^{-1}$ Hz$^{-1}$) versus timescale phase space of radio transients at frequencies of a few GHz.
The late-time afterglows of NS-NS/NS-BH mergers (red ellipses) are one the best electromagnetic counterparts expected for aLIGO sources \cite{hotokezaka2016}.
They are distinguished from other transient classes (black ellipses; shaded region represents AGN) in this phase space.
Some overlap exists with Type II supernovae, but the latter are expected to be optically thick at peak, while the former will be optically thin. 
Filled circles, open triangles and squares represent on-axis GRBs \citep{chandra2012}, Type II and Type Ib/c SNe \citep{weiler2002,stockdale2006,soderberg2010} respectively followed up in the radio. 
Additionally, the low transient rates imply low contamination by false positives. 
Note that the (early-time) radio afterglow of GW\,170817 has a luminosity of $\sim$10$^{26}$ and a timescale of $>1$ month \citep{radio-gw170817,mooley2017-gw170817}.
The right y-axis denotes the detectability distance for a deeper survey than ours, having a detection threshold of 300 $\mu$Jy.
Figure adapted from \cite{frail2012} and \cite{mooley2016}.}
\label{fig:lum_tau}
\end{figure}

We have used the VLA to image a 100 deg$^{2}$ error region of the O1 aLIGO BBH merger event GW\,151226.  
While a bright EM counterpart was not expected in this case, we have demonstrated a near real-time ability to conduct blind searches in a phase space where the predicted afterglow signature is strongest. 
As we noted in \S\ref{sec:intro}, a late-time radio afterglow peaking on timescales of 100's of days is one of the more robustly predicted afterglow signatures from neutron star mergers \citep{nakar2011}. 
Such searches are best carried out at GHz frequencies, since synchrotron self-absorption suppresses the signal below 1 GHz for up to several years post-merger. 
Estimates of the radio afterglow signal currently suffer from uncertainties in the circum-merger gas density, and a full search must be sensitive to the low densities (10$^{-3}$ cm$^{-3}$), as seen for 
some short gamma-ray bursts \citep{fbm+14}, and the density within galactic disks, $\sim$1 cm$^{-3}$ \citep{draine11}, as expected for the Galactic binary pulsar population.

\cite{metzger2012} have listed the ''cardinal virtues'' that must be met for an experiment to be considered a competitive follow-up effort for detecting the predicted signatures of compact binary star mergers. 
These are 1) they are detectable with current observing facilities with reasonable time allocation,  2) they accompany a significant fraction of GW events, 3) they can be distinguished from other astrophysical transients, 
and they can provide arcsecond sky localization. 
Below we demonstrate how this experiment addresses the challenges and meets almost all of these requirements.

Since the field-of-view of the VLA at 3 GHz is small (0.06 deg$^2$), we take advantage of the superb instantaneous sensitivity of the VLA and use the OTFM method, rapidly slewing the antennas (1.6$^\circ$/min) in order 
to image a 100 deg$^{2}$ error region with only 7.5 hrs of on-sky integration time.  
These observations were made at three epochs approximately 50, 101 and 185 days post-merger (see Table~\ref{tab:observations}). 
These are an appropriate range of timescales for searching for the rise and fall of the expected late-time radio afterglow. 
The typical RMS noise achieved for each epoch was approximately 120 $\mu$Jy/beam. 
This is an astrophysically interesting sensitivity level for detecting afterglows circum-merger gas densities $>$0.1 cm$^{-3}$ and distances of 100--200 Mpc \citep{hotokezaka2016}.

We searched for any transient sources that appeared in the 100 deg$^{2}$ region over the three epochs. 
None were found. 
This implies an upper limit to the areal density of transients brighter than 700 $\mu$Jy of $10^{-2}$ deg$^{-2}$. 
Our limit is similar to that obtained from the 3 GHz Stripe 82 Pilot project \citep{mooley2016}. 
The types and timescales of the different radio transients at frequencies of a few GHz are shown in Figure~\ref{fig:lum_tau}. 
On the timescales that were sampled in this experiment, the expected transients include tidal disruption events (TDE), orphan GRB afterglows, Type II core-collapse supernovae.
If the gravitational wave event is contained within our survey region, we can place an upper limit of about $10^{29}$ erg s$^{-1}$ Hz$^{-1}$ to any late-time radio afterglow from the merger event at an assumed mean distance of 440 Mpc. 
At an areal density of $10^{-2}$ deg$^{-2}$, none of these transient source classes are expected to be a significant contaminant in a blind radio survey for the late-time afterglow from a binary compact remnant merger \cite{mooley2016}. 
The quietness of the radio sky stands in contrast to optical counterpart searches \citep[e.g][]{cbs+16,csp+16,pck+16}, for which there were a number of transients that required spectroscopic 
follow-up before being eliminated as candidates. 
Radio observations of GW sources also have an advantage over the optical and X-ray in cases where solar/lunar observing constraints are present and dust-obscured environments 
in the host galaxy itself, or along lines of sight through our own Galaxy. 
For example, continued radio monitoring of GW\,170817 was carried out during a crucial period in the evolution of the afterglow light curve, while optical and X-ray telescopes were constrained by the Sun, 
thus allowing us to distinguish between the off-axis jet versus cocoon models \citep{mooley2017-gw170817}.

\begin{figure}
\centering
\includegraphics[width=3.5in]{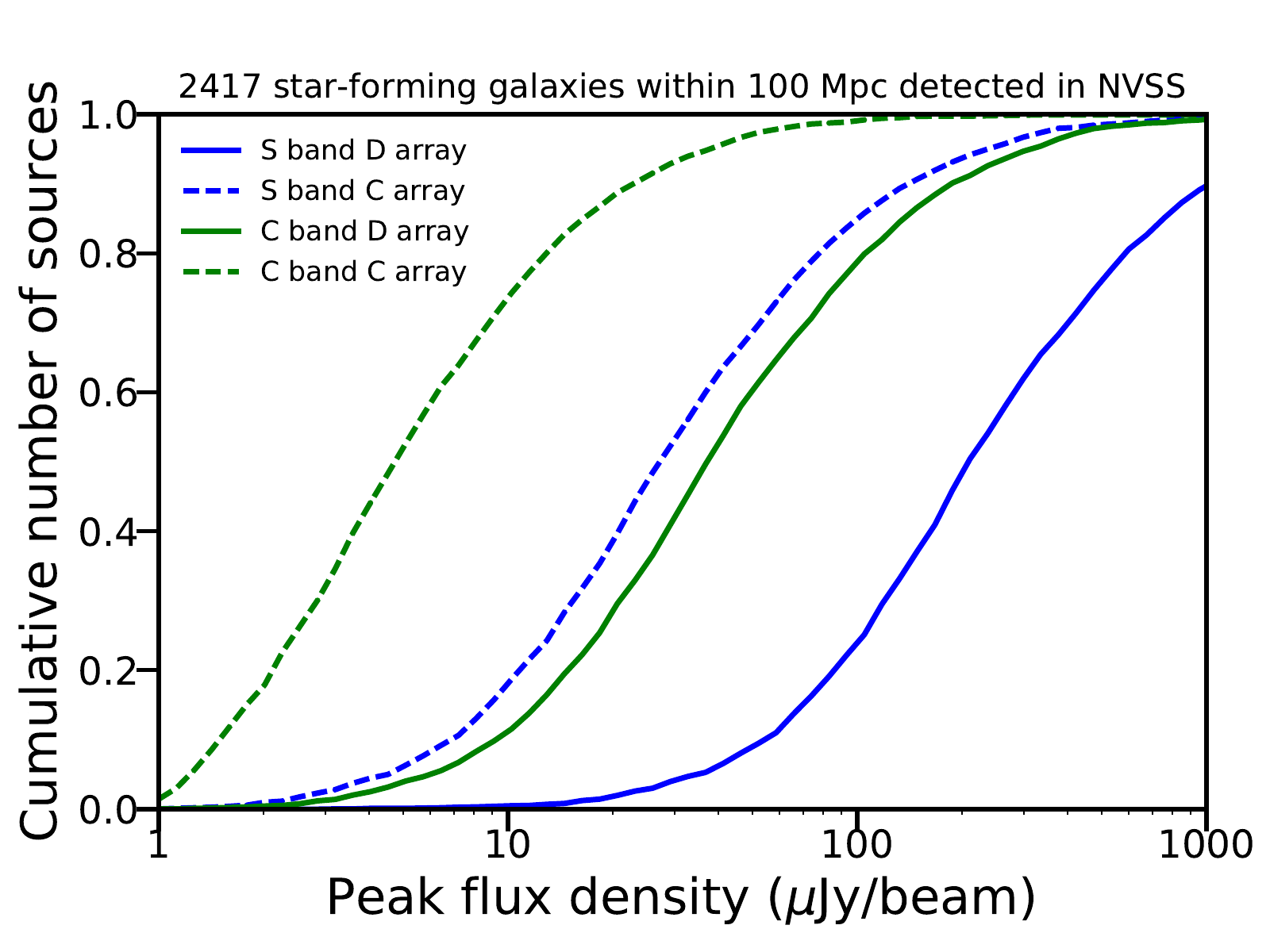}
\caption{The cumulative plots for the peak flux densities of star-forming galaxies within 100 Mpc \citep[detected in the NVSS;][]{condon2002} at S and C bands 
(3 GHz and 6 GHz respectively) for the VLA C and D array configurations.
Contamination by the host galaxy may be a significant problem for merger afterglows in D config, but searches at C band will abate this.
Beyond 100--150 Mpc, contamination becomes less severe \citep{hotokezaka2016}.
If the merger afterglows reside within the galactic disks, they will make the host galaxies appear as variable sources.
Two fiducial surveys with source detection thresholds of 100 $\mu$Jy and 300 $\mu$Jy will be able to reliably recover variable point sources in 
host galaxies having peak flux densities less than $\sim$400 $\mu$Jy and $\sim$1200 $\mu$Jy (i.e. corresponding to 25\% variability).
This corresponds to 70\% (95\%) and 90\% (100\%) of the host galaxies respectively within 100 Mpc at S band (C band) in D config.
An alternative, but more challenging, method will be to employ host-galaxy subtraction.
In this plot, correction for the observing frequency has been made assuming spectral index of $-0.7$.
Integrated flux density and size of the radio source have been converted to the peak flux density using the equation for face-on galaxy from \cite{condon2015}.}
\label{fig:starformation_contamination}
\end{figure}

\begin{figure}
\centering
\includegraphics[width=3.5in]{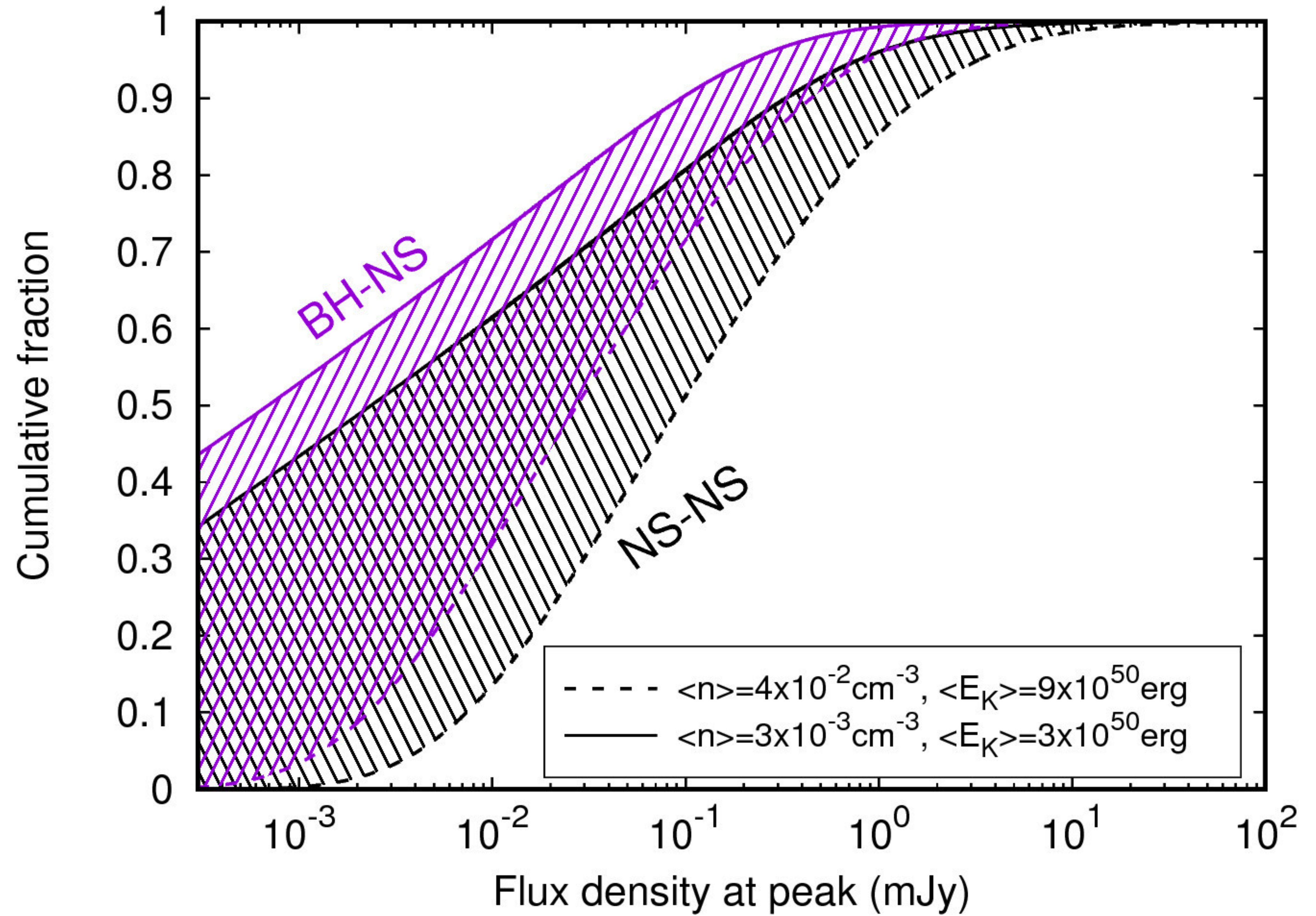}
\caption{Cumulative flux density distributions expected for double neutron star (NS-NS; black region) and black hole-neutron star (BH-NS; purple region) mergers 
(those detected as gravitational wave sources) at the peaks of their light curves.
The distributions suggest that blind radio searches with RMS noise values of 50--75 $\mu$Jy/beam will be able to access up to 40\% of the merger afterglows.
For calculating these curves we use the density distribution derived from short-GRB afterglow observations \citep[similar to the $\nu_c<\nu_x$ case for an optimistic choice 
and the $n_{\rm min}=10^{-6}$\,cm$^{-3}$ case for a pessimistic choice; ][]{fong2015}.  
We use a hypothetical log-normal distribution for the kinetic energy with a peak at $9 \times 10^{50}\,(3\times10^{50})$\,erg for an optimistic (pessimistic) choice and the standard 
deviation chosen to be unity. 
The distribution of the distances of binary neutron and black-hole neutron star mergers is taken from \cite{hotokezaka2016}.
The microphysical parameters are set to: $\epsilon_e=\epsilon_B=0.1$ and $p=2.5$.
Note that the only difference between the NS-NS and BH-NS cases (as plotted) is the distance distribution; the latter occur at farther distances so their flux densities are lower.
However, the kinetic energies involved in BH-NS mergers may be larger, depending on the black hole spins, mass ratios and neutron star equation of state. }
\label{fig:cum_flux}
\end{figure}

While no transients were found, about 1.5\% of the compact, persistent radio sources in the field exhibited significant variability. 
Most radio sources vary to some degree or another. 
Below $\sim$1 GHz the variations are dominated by plasma effects such as interstellar scintillation, while at high frequencies the variations are intrinsic, likely originating 
from changes in the black hole-disk-jet environments \citep{altschuler1989}. 
The frequency and level of variability is in agreement with previous GHz studies \citep{mooley2013,mooley2016,hdb+16} and is consistent with normal AGN 
activity\footnote{We have manually inspected the optical/infrared images and photometric colors of the variable sources having multiwavelength counterparts and 
cross-matched with AGN catalogs to confirm that these are indeed AGN.} on a timescale of a few months.
In principle, a transient occurring in a star-forming galaxy could be mistakenly identified as an unresolved or partially-resolved variable source. 
In practice, with the 8 arcsecond synthesized beam and the RMS noise of $\sim$120 $\mu$Jy/beam, the experiment does not have sufficient surface brightness sensitivity 
to detect star-forming galaxies within the LIGO volume \citep{condon2015} for GW\,151226. 
However, host galaxy contamination may be severe in the cases where merger afterglows are located within $\sim$100 Mpc and observations are carried out 
in VLA D config (see Figure~\ref{fig:starformation_contamination}).
\cite{hotokezaka2016} looked at contamination as a function of galaxy distance for VLA B config and L band (1.4 GHz), and they find that 
contamination due to star-burst emission and bright AGN will be $\sim$5\%.
 Nevertheless, an interferometer such as the VLA not only provides a sub-arcsecond position for a potential merger event, 
but it also acts as a spatial filter of extended host galaxy emission.

While we have shown that GHz radio searches for late afterglows are a promising method to search for the EM signature of a GW event, 
the experiment could still be improved. 
In \S\ref{sec:obs_proc.cal} we noted that 1.8 days were required to process each epoch. 
Lower latency would be possible if the data calibration stage was further optimized. 
The benefits of such improvements are not immediately obvious. 
The radio afterglow reaches a peak months after the gravitational event, well after any putative optical signature has faded away. 
The follow up of any radio candidates would be with a goal of identifying a host galaxy and measuring its redshift, and as such could be done at a more leisurely pace.

A weakness of radio searches is that the large uncertainty in the expected circum-merger densities predicts a wide range of peak flux densities at GHz frequencies (see \S\ref{sec:intro}). 
Figure~\ref{fig:cum_flux} shows the cumulative distributions of peak flux densities of neutron star merger \citep[simulated population of gravitational wave sources;][]{hotokezaka2016}.
It is possible to achieve noise values of 50--75 $\mu$Jy/beam and thus broaden the parameter space of the late-time afterglow models. 
Such deeper searches will be able to access up to 40\% of the merger afterglows.
For this blind survey approach and with finite VLA telescope time, the trade off is to image smaller fields of view. 
Such decreases in the size of the GW error regions are expected in the upcoming runs jointly with LIGO and Virgo, and eventually runs including KAGRA and LIGO-India \citep{nkg13,kn14,abbott2016}. 
Another approach for increasing sensitivity is to significantly reduce the total area searched by doing targeted searches. 
This approach involves deep radio follow-up of transient candidates identified at other wavelengths \citep{pck+16,bhalerao-atlas} or using 
a master galaxy catalog in the localized volume \citep{gck16,sch+16}. 
Such a strategy has been successful in the case of GW170817.
However, as the distance range for the gravitational wave detectors increase, blind surveys or even hybrid approaches between targeted and blind searches will become important.

\acknowledgments{\it The National Radio Astronomy Observatory is a facility of the National Science Foundation operated under cooperative agreement 
by Associated Universities, Inc.
K.P.M acknowledges support from the Oxford Centre for Astrophysical Surveys which is funded through the Hintze Family Charitable Foundation.
K.P.M. is currently a Jansky Fellow of the National Radio Astronomy Observatory.
We thank the anonymous referee for providing comments that helped improve the manuscript.}


\end{document}